\documentclass[]{spie}  
\usepackage[dvips]{graphicx}

\title{Humidity contribution to the refractive index structure function $C_n^2$.}

\author{Carlos O. Font, Mark P. J. L. Chang\supit{1}, Eun Oh and Charmaine Gilbreath\supit{2}
\skiplinehalf
\supit{1}Physics Department, University of Puerto Rico, Mayag\"uez, Puerto Rico 00680 \\
\supit{2}U.S. Naval Research Laboratory, Washington D.C. 20375
}

\authorinfo{Further author information: (Send correspondence to M.P.J.L.C.)\\M.P.J.L.C.: E-mail: mchang@uprm.edu, Telephone: 1 787 265 3844}
 
\begin{document} 
  \maketitle 

\begin{abstract}
Humidity and $C_n^2$ data collected from the Chesapeake Bay area during the 2003/2004 period have been analyzed.  We demonstrate that there is an unequivocal correlation between the data during the same time periods, in the absence of solar insolation.  This correlation manifests itself as an inverse relationship.  We suggest that $C_n^2$ in the infrared region is also function of humidity, in addition to temperature and pressure.
\end{abstract}



\section{INTRODUCTION}
\label{sect:intro}

It has been known for some time\cite{Roddier1981} that the
scintillation behaviour of point sources is a measure of the optical seeing in
the atmosphere.  What has been less well understood is the contribution of
different environmental variables to optical seeing.  Over the past decade,
a great deal of study has been dedicated to clarifying this issue.

Comprehensive treatments of the theory of wave propagation in random media are
given in Tatarskii's seminal works\cite{Tatarskii1961,Tatarskii1971}.  More recent
developments are described in Tatarskii et al.\cite{Tatarskii1992}.  Some of the
simplest models based on these complex works are well known and available in
the literature: Greenwood\cite{Greenwood1977},
Hufnagel--Valley\cite{HufnagelValley1974}, 
SLC-Day and SLC-Night\cite{Miller1976}.
These models are used to predict the strength of weak clear air
turbulence's refractive index structure function, $C_n^2$, but in all cases
they have major failings: either they are too general and do not take into
account local geography and environment (as in the former two) or they are too
specific to a site (as in the latter two, which reference the
median values above Mt. Haleakala in Maui, Hawaii).

A more recent numerical model known as PAMELA does attempt to
account for geographical position and ambient climate factors.  
However, its inverse power windspeed term fails to explain some of the characteristics of $C_n^2$ during low wind conditions.  It does an adequate job for characterizing horizontal and diagonal beam propagation within the atmospheric boundary layer\cite{Han2004}.

Despite the differences, the models agree in terms of the overall general 
behaviour of $C_n^2$.  For example, it is to be expected that during the
daylight hours, the $C_n^2$ trend will be dominated by the solar
insolation and in those models that do account for day/night differences this is presented. The physical effect is evidenced in Oh\cite{Oh2004}, where in many
cases scintillometer measurements are seen to strongly follow the measured
solar insolation function.  When the sun sets however, it is less clear as to
the predominant contributing factors.  In an extension of earlier work,
Oh presented indications of a possible anticorrelation effect between the ambient relative humidity and the value of $C_n^2$.

In this paper, we report on further analysis of the datasets obtained during
that study to show that there is an unequivocal correlation in the absence of solar insolation in a littoral space.

\section{INSTRUMENTS AND ALGORITHMS} 
\label{sect:expt}

The $C_n^2$ and associated weather variable data was collected over a
number of days during 2003 and 2004 at the Chesapeake Bay Detachment (CBD) of
the Naval Research Laboratory.

The $C_n^2$ data was obtained with a
commercially available scintillometer (model LOA-004) from Optical Scientific
Inc, which serves as both a scintillometer and as an optical anemometer for
winds transverse to the beam paths.  The local weather parameters were
determined by a Davis Provantage Plus (DP+) weather station.  The LOA-004 had a
sample rate of 10 seconds, while the DP+ was set at 5 minutes.

The LOA-004 instrument comprises of a single modulated infrared transmitter
whose output is detected by two single pixel detectors.  For these data, the
separation between transmitter and receiver was 100-m.
The path integrated $C_n^2$ measurements are determined by the LOA instrument
by computation from the log--amplitude scintillation ($C_\chi(r)$) of the two
receiving signals\cite{Ochs1979,Wang2002}.  The algorithm for relating $C_\chi(r)$ to $C_n^2$ is based on an equation for the log--amplitude covariance function in Kolmogorov turbulence by Clifford {\em{et al.}}\cite{Clifford1974}, which we repeat here
\begin{equation}
C_\chi(r) = 2.94 \int^1_0 du \sigma_T^2(u)[u(1-u)^{5/6}] \int_0^\infty dy y^{-11/6} sin^y \exp \{ -\sigma^2_T [u (1-u) ]^{5/6} F(y) \} J_0 \left\{ \left[ \frac{4 \pi y u}{(1 - u)} \right]^{1/2} r \right\}
\end{equation}
The terms in this equation are: $r$, the separation between two point detectors in Fresnel zones $\sqrt{\lambda L}$, with $L$ being the path distance between source and detectors; $y$ is the normalized spatial wavenumber; $u = z/L$ is the normalized path position; $J_0$ is the zero order Bessel function of the first kind and 
\begin{eqnarray}
\sigma_T^2 (u) & = & 0.124 k^{7/6} L^{11/6} C_n^2(u) \\ \nonumber
F(y) & = & 7.02 y^{5/6} \int_{0.7y}^\infty d\xi \xi^{-8/3} [1 - J_0(\xi)]
\end{eqnarray}
This can be better appreciated if we define a path weighting function $W(u)$ such that
\begin{equation}
C_\chi(r) = \int^1_0 du C_n^2 (u) W(u)
\end{equation}
for a point source and point receivers where
\begin{equation}
W(u) = 0.365 k^{7/6} L^{11/6} [u(1-u)]^{5/6} \int^\infty_0 dy g(u,y) J_0 \left\{ \left[ \frac{4 \pi y u}{(1-u)} \right]^{1/2} r \right\}
\end{equation}
In the above expression, $g(u,y)$ carries the information related to $C_n^2$ for point source and point receivers.  It can be modified to incorporate finite receiver and transmitter geometries.

Some comments are necessary at this point.  The key assumptions made by the
LOA-004 instrument in computing $C_n^2$ are:
\begin{itemize}
\item The turbulent power spectrum is Kolmogorov; the spatial power
  spectrum of temperature fluctuations $\Phi_T^2 (k)$ and the humidity
  fluctuations $\Phi_H^2 (k)$ are proportional to $k^{-5/3}$.  This may not
  always be true, especially if the inner and outer scales are on the order of
  the relevant dimensions of the observing system.
\item As a result of the previous assumption, the index of refraction
  structure function is assumed to be dependent only on the 
  temperature structure function and pressure at optical frequencies.
\end{itemize}
What we demonstrate in this paper is that the LOA-004 measured $C_n^2$ function from the CBD experiment is indeed correlated with the humidity.  The LOA-004's design is by no means optimal for extracting $C_n^2$, since its main purpose is to act as an anemometer.  To that end we have an effort to evaluate contributions to the measurement error of $C_n^2$.

\section{ANALYSIS} 
\label{sec:analysis}

\begin{figure}
\begin{center}
\begin{tabular}{c}
\includegraphics[height=0.4\textwidth]{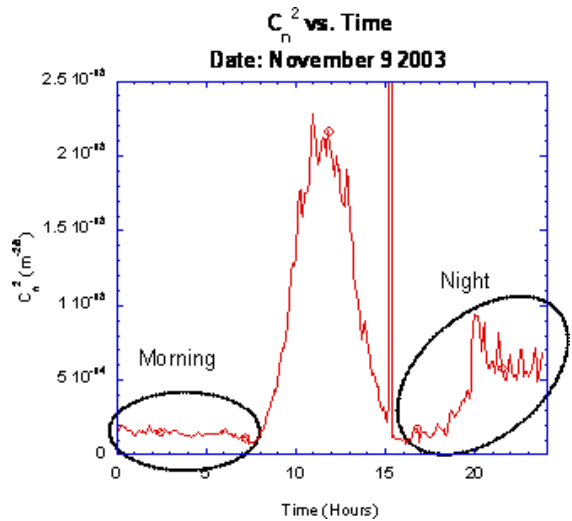}
\end{tabular}
\end{center}
\caption{ \label{fig:DayNight} Morning and Night definitions for our purposes.  See text for details. }
\end{figure}

The $C_n^2$ data was smoothed with a 60 point rolling average function.  The effect of solar insolation was excluded from this study.  We defined the morning and night portions of a 24 hour period as shown in Figure (\ref{fig:DayNight}).
Morning runs from midnight until sunrise (as corroborated by a solar irradiance measurement), while night runs from sunset to 23:59.  As reported in Oh {\em et al.}\cite{Oh2004} visual inspection of the valid time series data gives the impression that there is an approximate inverse relationship between $C_n^2$ and humidity.  This can be appreciated in a more quantitative manner by graphing $C_n^2$ against humidity.  

We chose data sets in which the temperature variations are no more than $\pm 15\%$  and the pressure change is at most 15 mbars over the time intervals of interest.  The data sections were also selected to have no scattering effects due to snow or rain, and the wind was northerly (to within approximately $\pm 20^\circ$, inflowing from the bay to land).

Given the aforementioned conditions, from the data available only a subset provided complete time series in both ambient weather variables and $C_n^2$. We were able to extract eight morning and evening runs, spanning seven days between November 2003 and March 2004 for the purpose of calculating the crosscorrelation, $\Gamma_{UV}(t+\delta t) = E[u(t+\delta t) v(t)]$, and cross covariance, $C_{UV} (t+\delta t) = E{(u(t+\delta t) - \overline{u})(v(t) - \overline{v})}$, between humidity and $C_n^2$ measureables.  In these parameters, $E$ represents expected value and $\overline{u},\overline{v}$ are the mean values of the two random processes, considered stationary.

\begin{figure}[!htp]
\begin{center}

\includegraphics[height=0.44\textwidth]{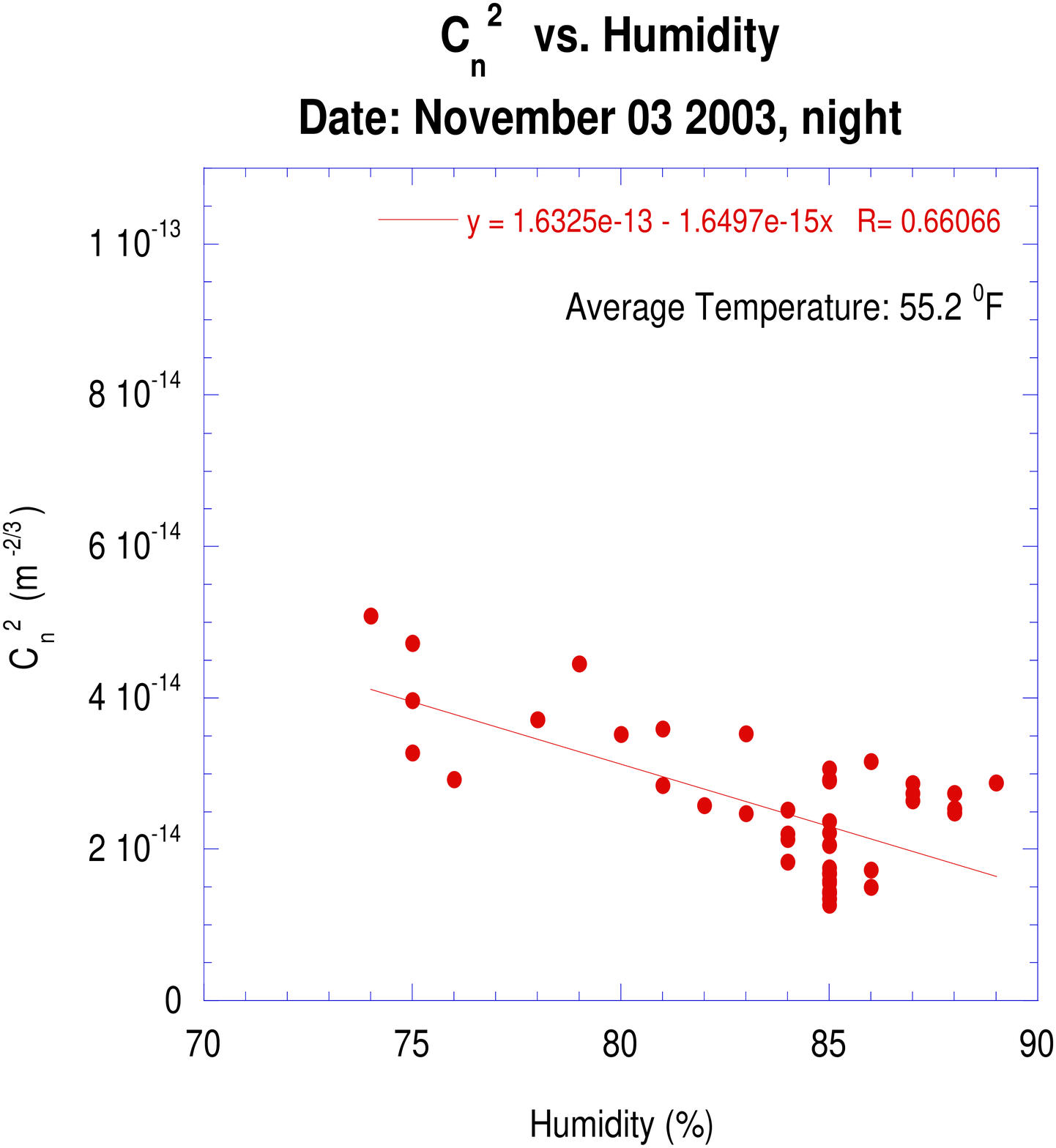}
\includegraphics[height=0.4\textwidth]{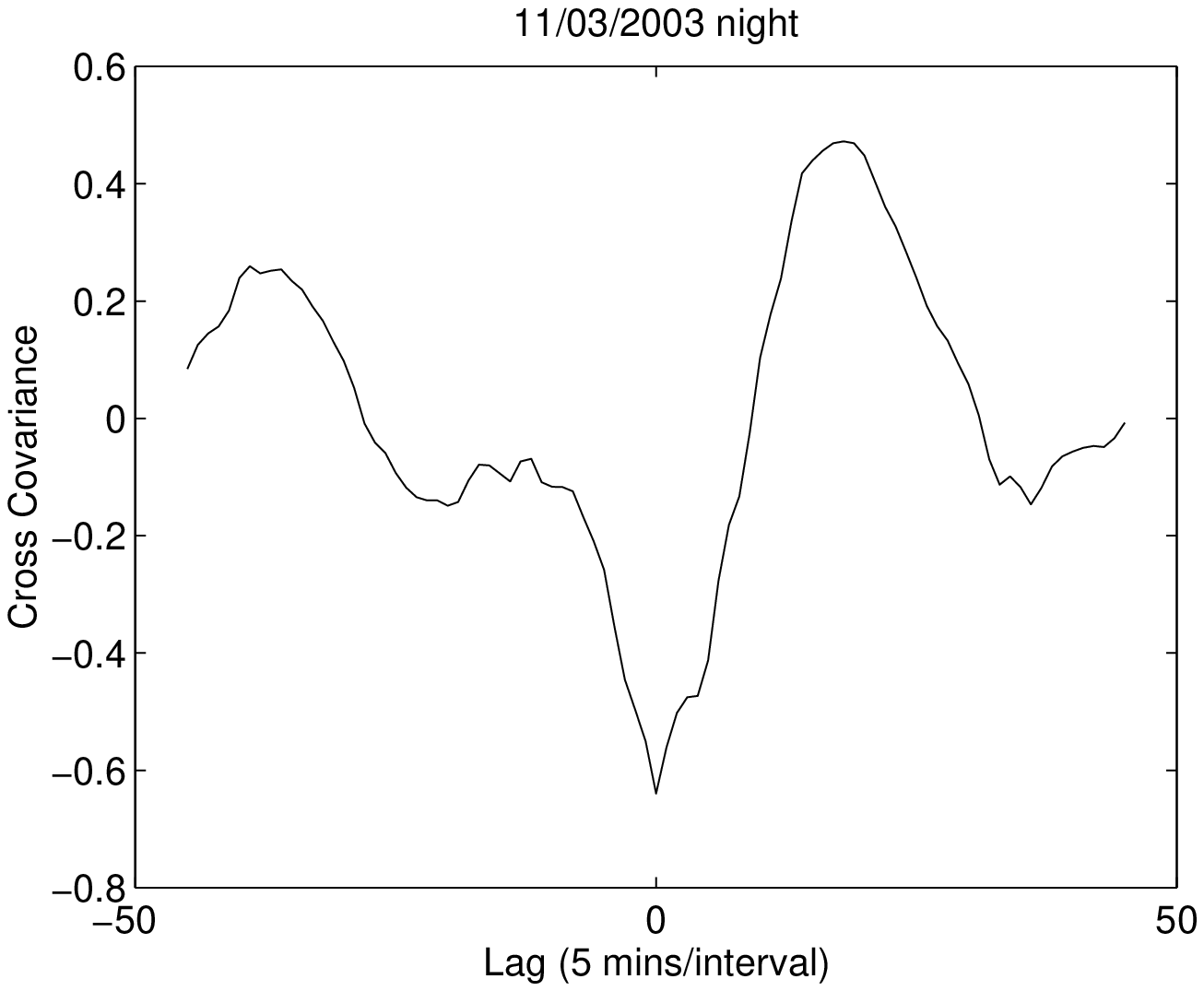}
\end{center}
\caption{\label{11032003} Nov 3 2003, Night}
\end{figure}

\begin{figure}[!htp]
\begin{center}
\includegraphics[height=0.44\textwidth]{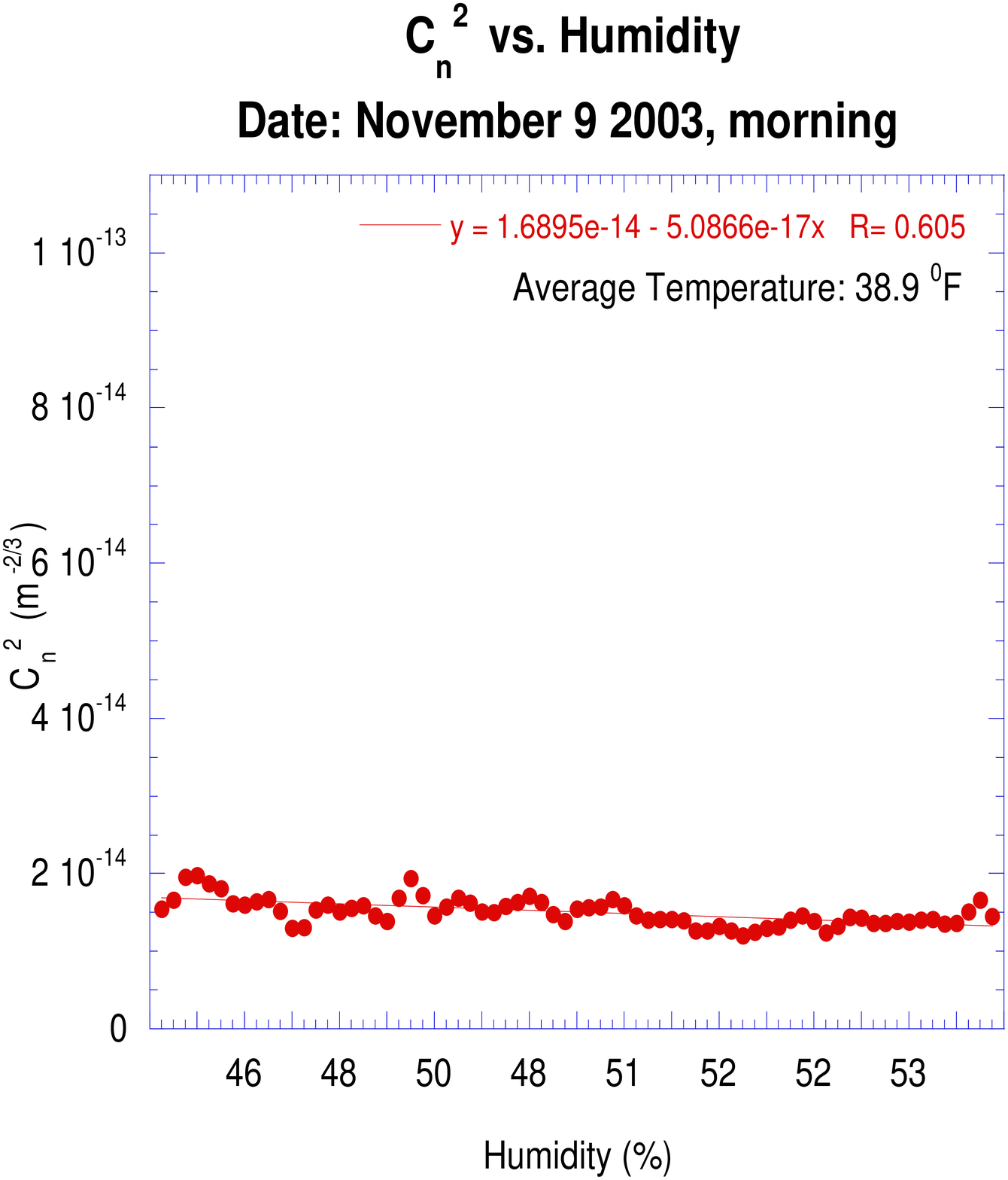}
\includegraphics[height=0.4\textwidth]{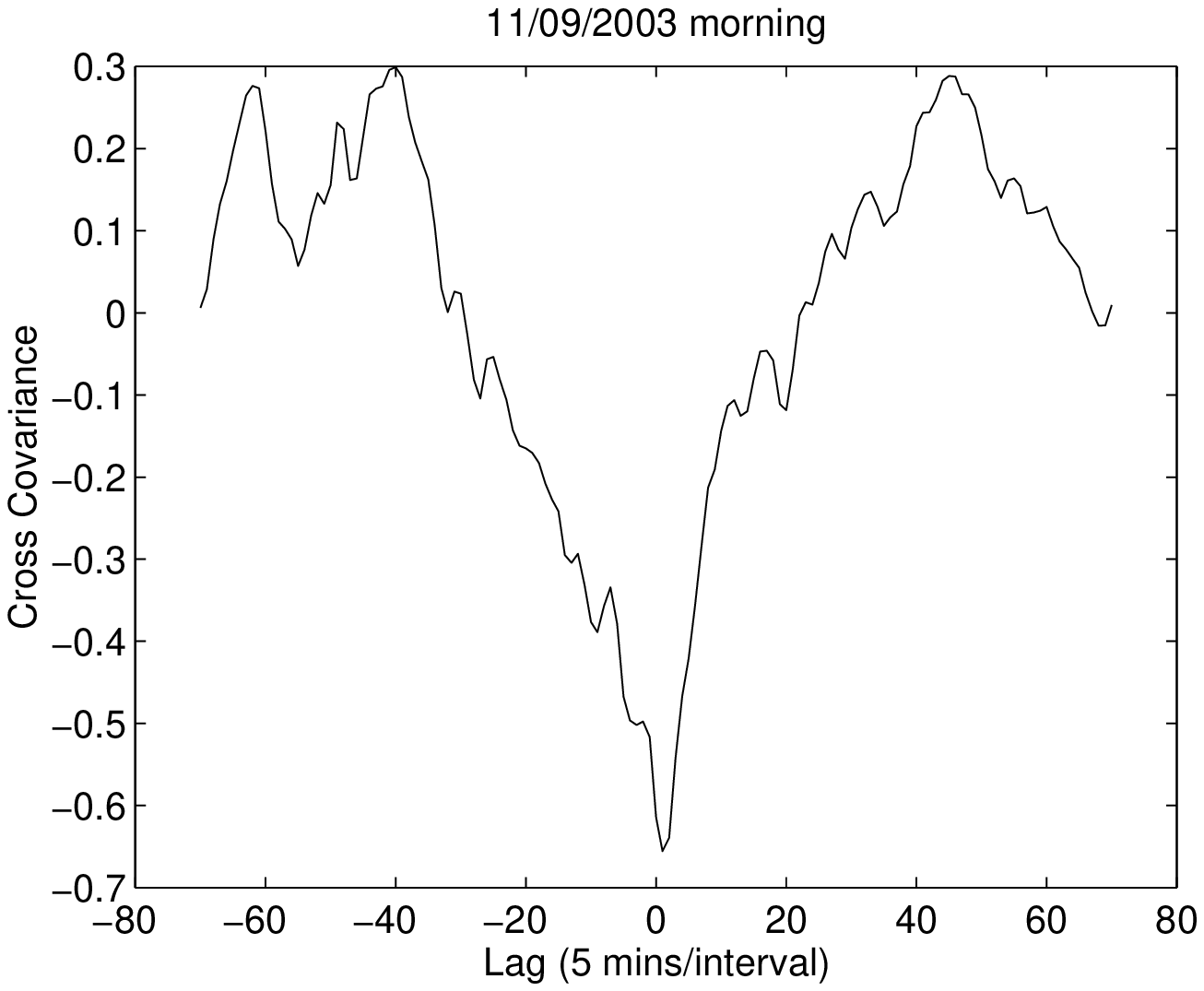}
\end{center}
\caption{\label{11092003} Nov 9 2003, Morning}
\end{figure}

\begin{figure}[!htp]
\begin{center}
\includegraphics[height=0.44\textwidth]{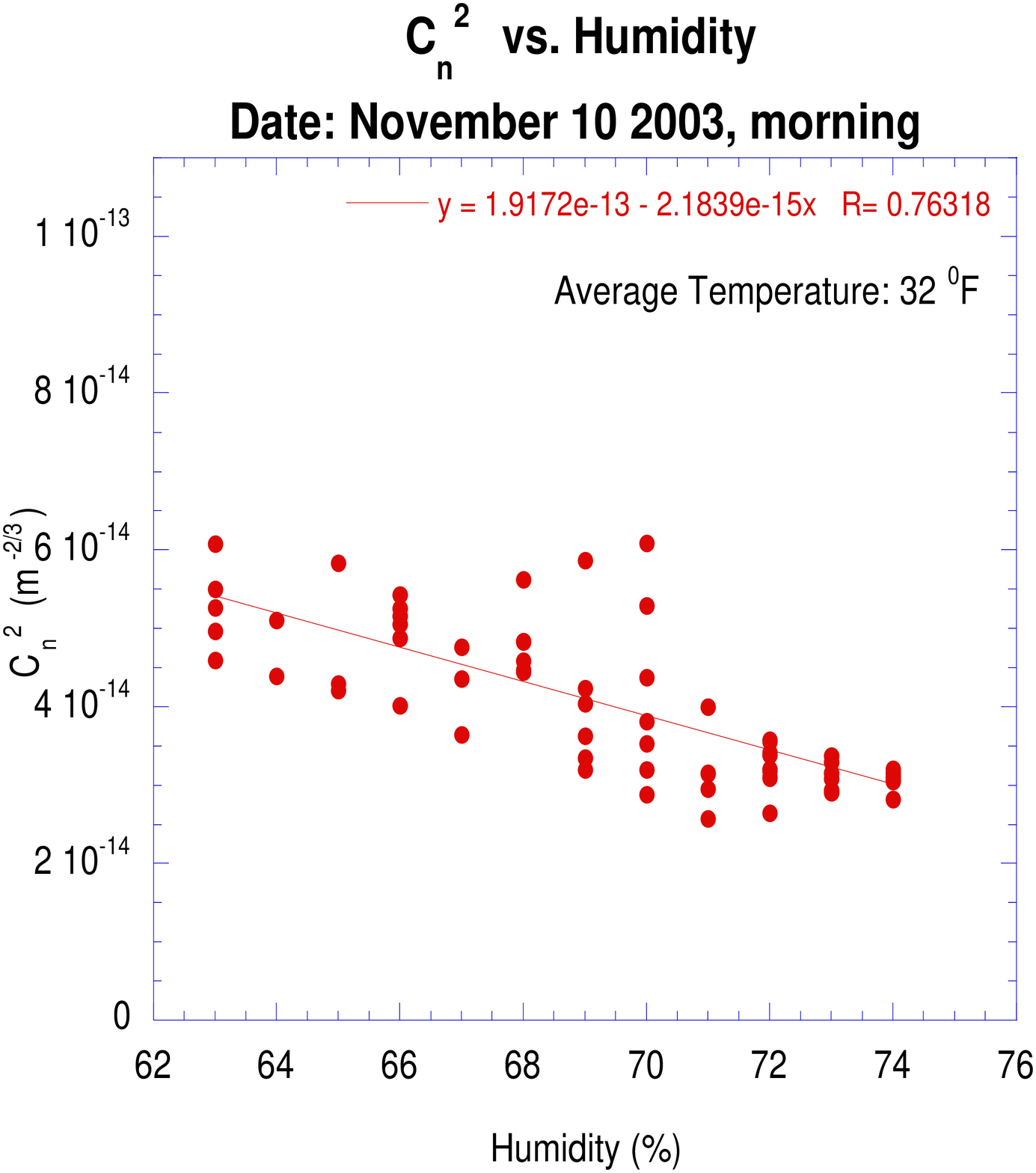}
\includegraphics[height=0.4\textwidth]{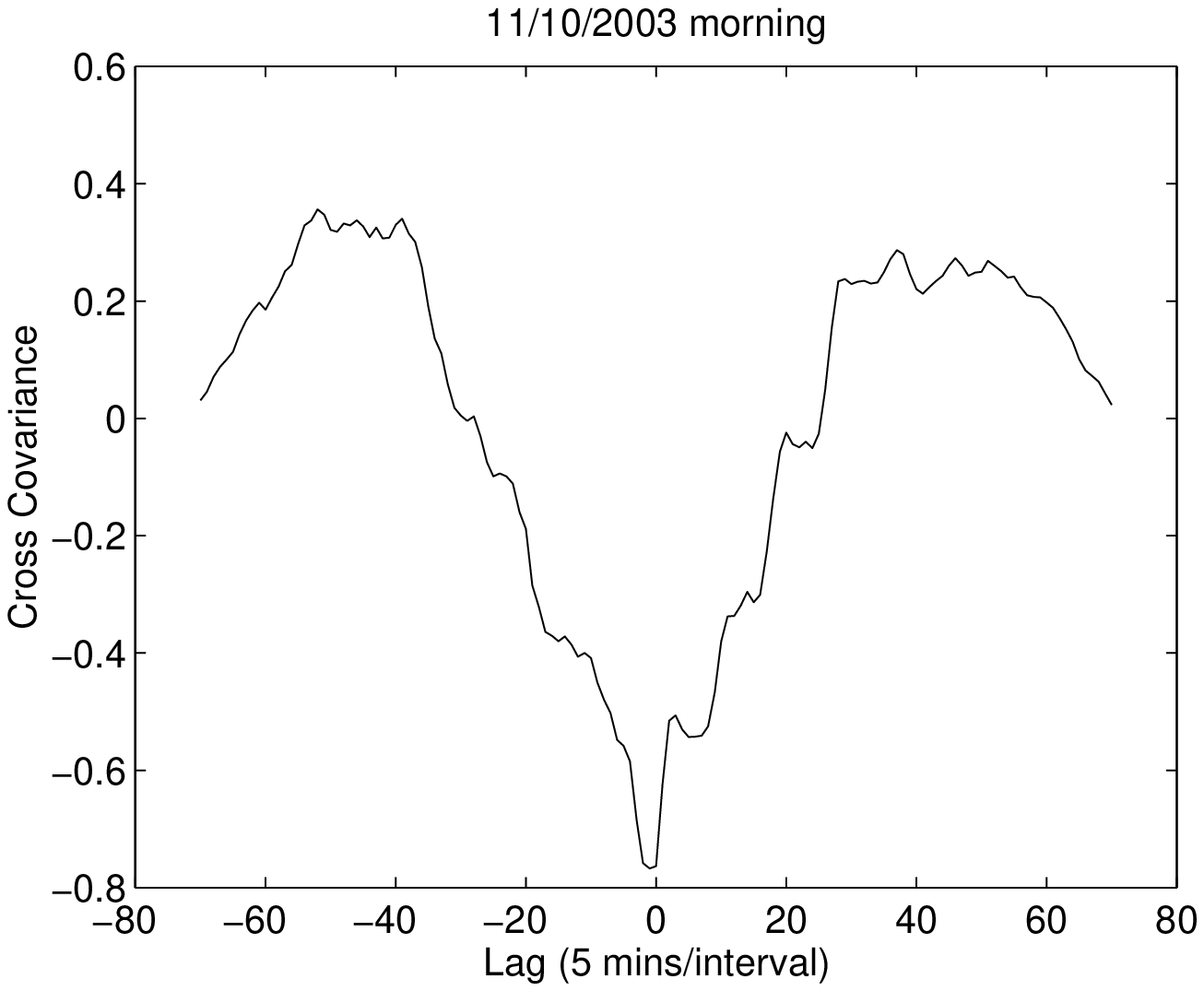}
\end{center}
\caption{\label{11102003} Nov 10 2003, Morning}
\end{figure}

\begin{figure}[!htp]
\begin{center}
\includegraphics[height=0.44\textwidth]{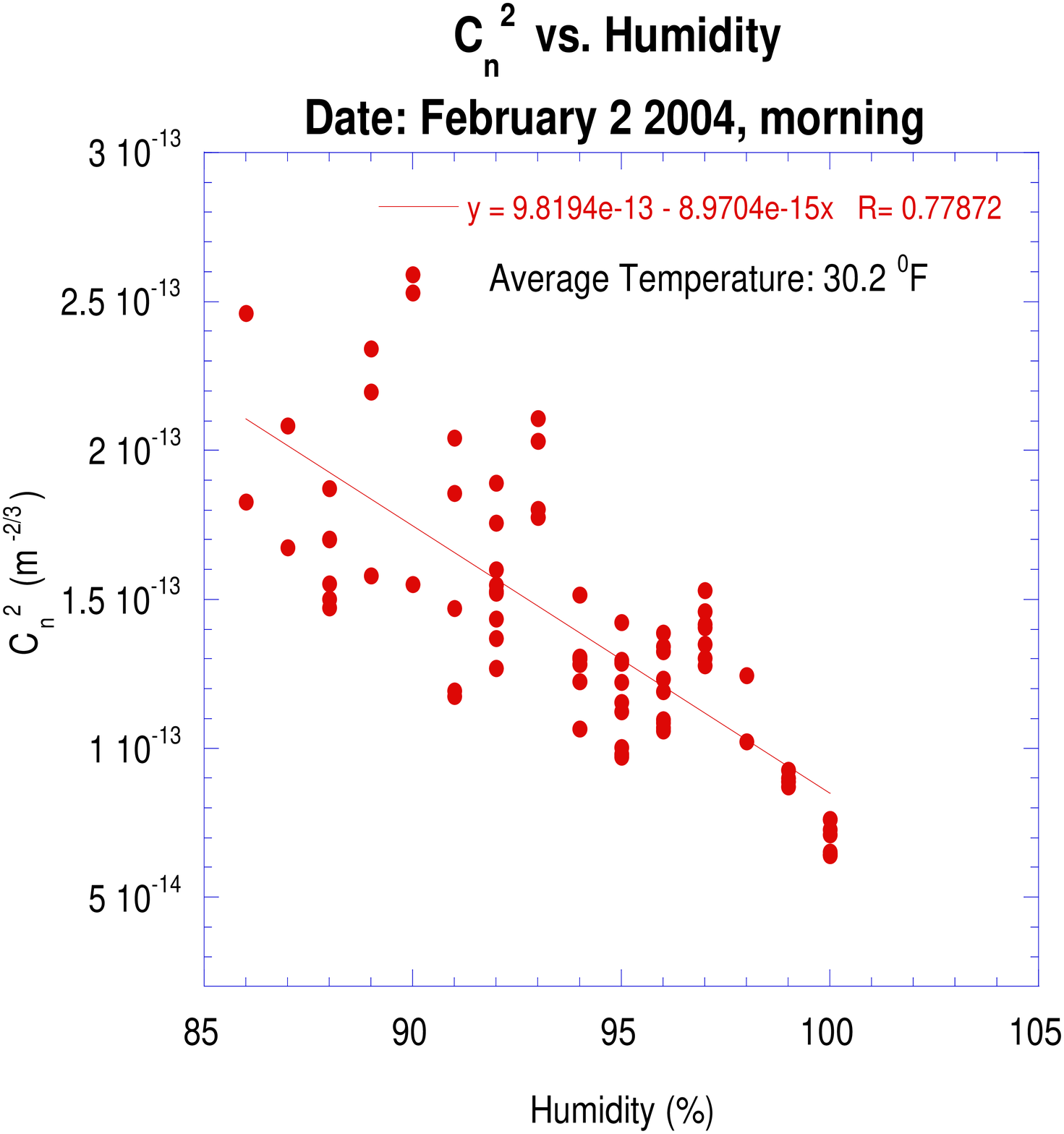}
\includegraphics[height=0.4\textwidth]{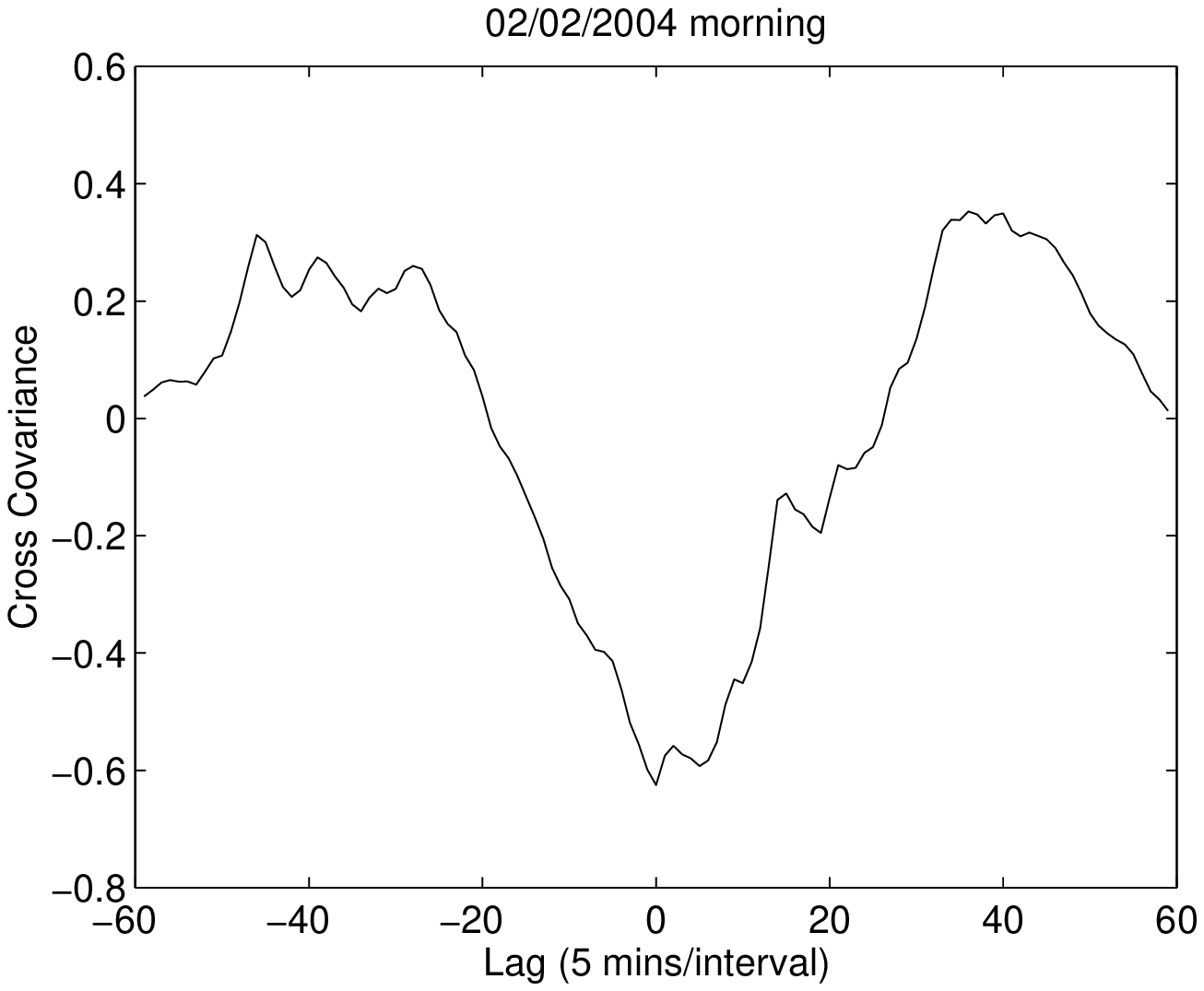}
\end{center}
\caption{\label{02022004} Feb 2 2004, Morning}
\end{figure}

\begin{figure}
\begin{center}
\includegraphics[height=0.44\textwidth]{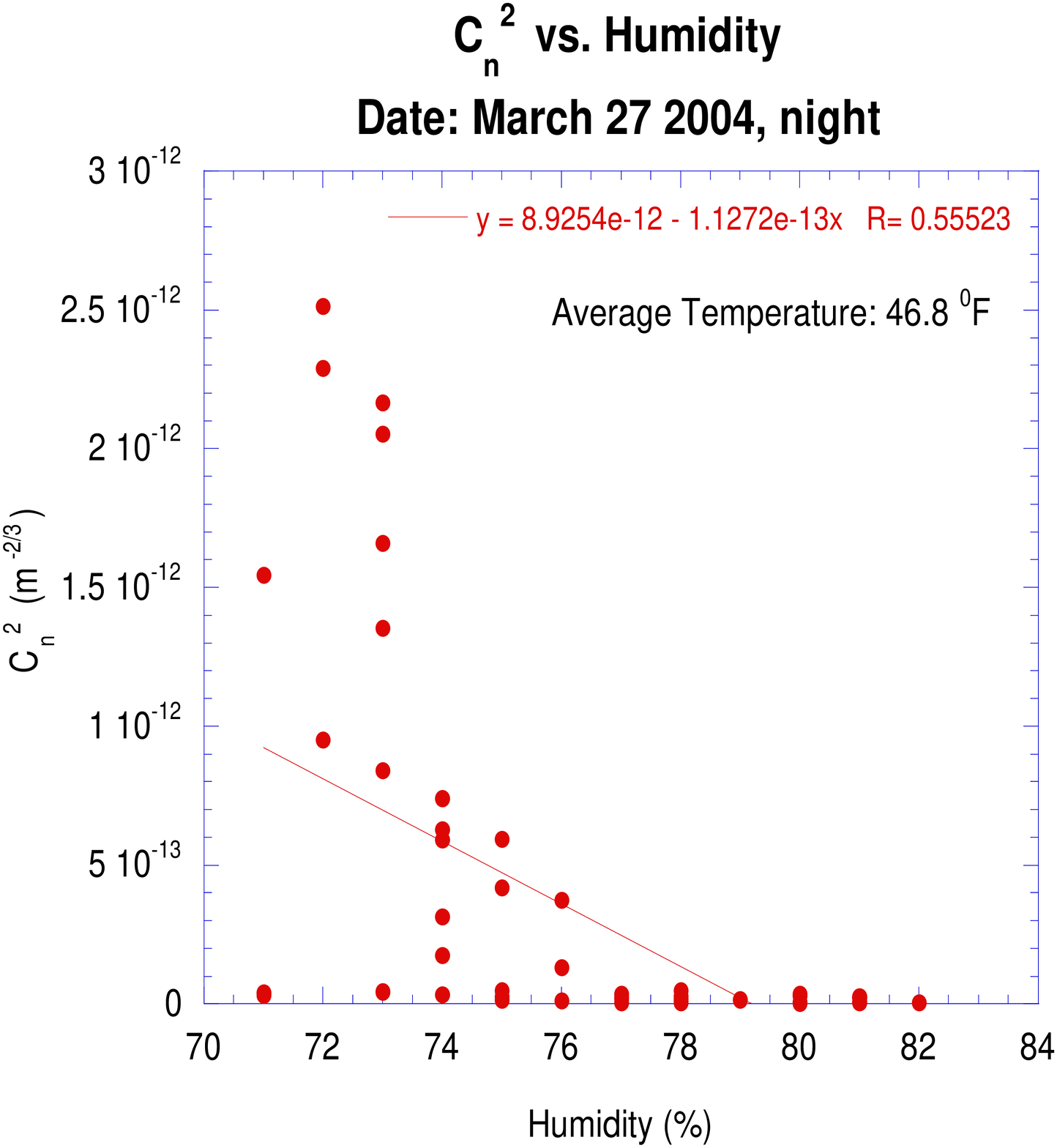}
\includegraphics[height=0.4\textwidth]{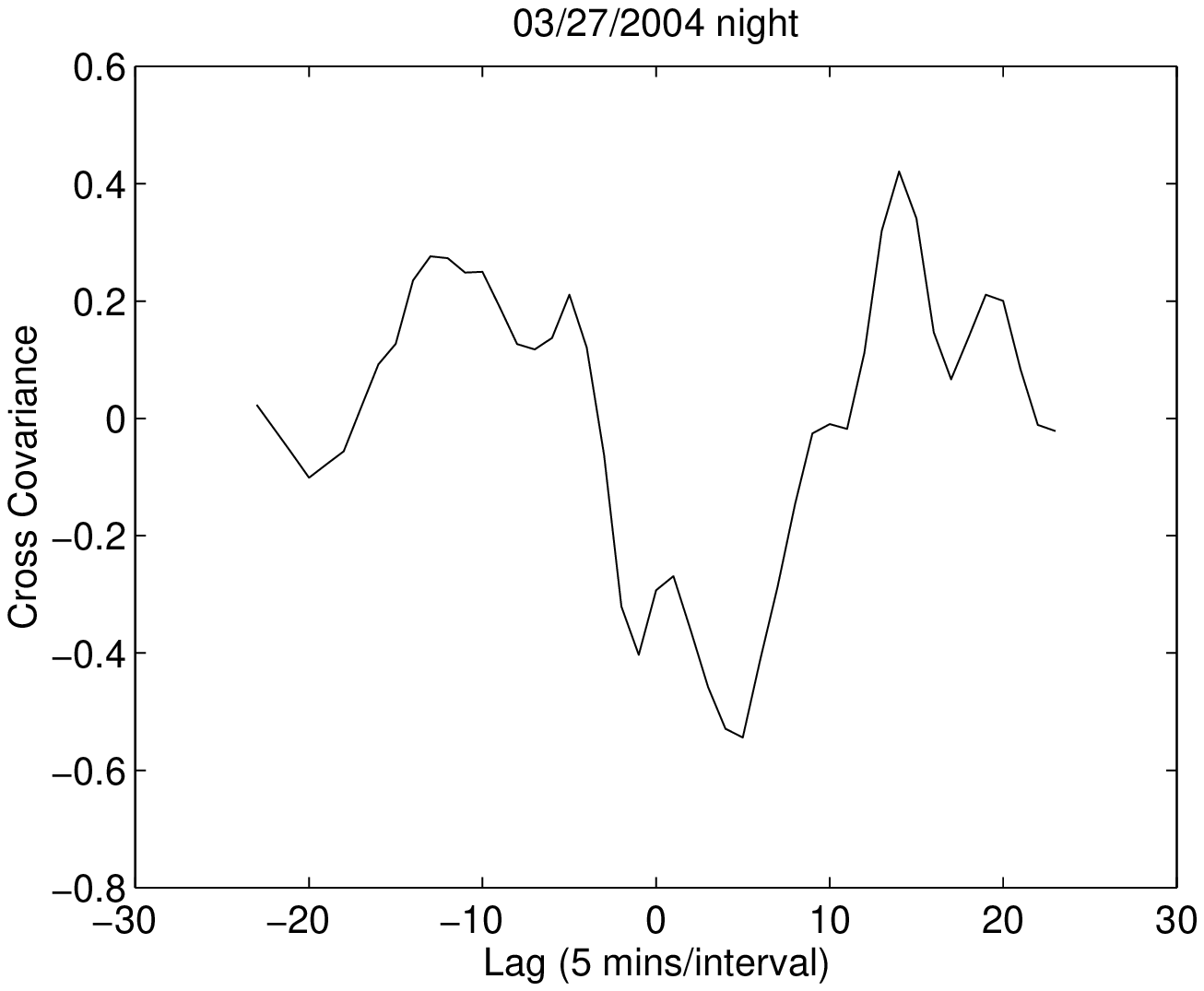}
\end{center}
\caption{\label{03272004} Mar 27 2004, Night}
\end{figure}

\begin{figure}
\begin{center}
\includegraphics[height=0.44\textwidth]{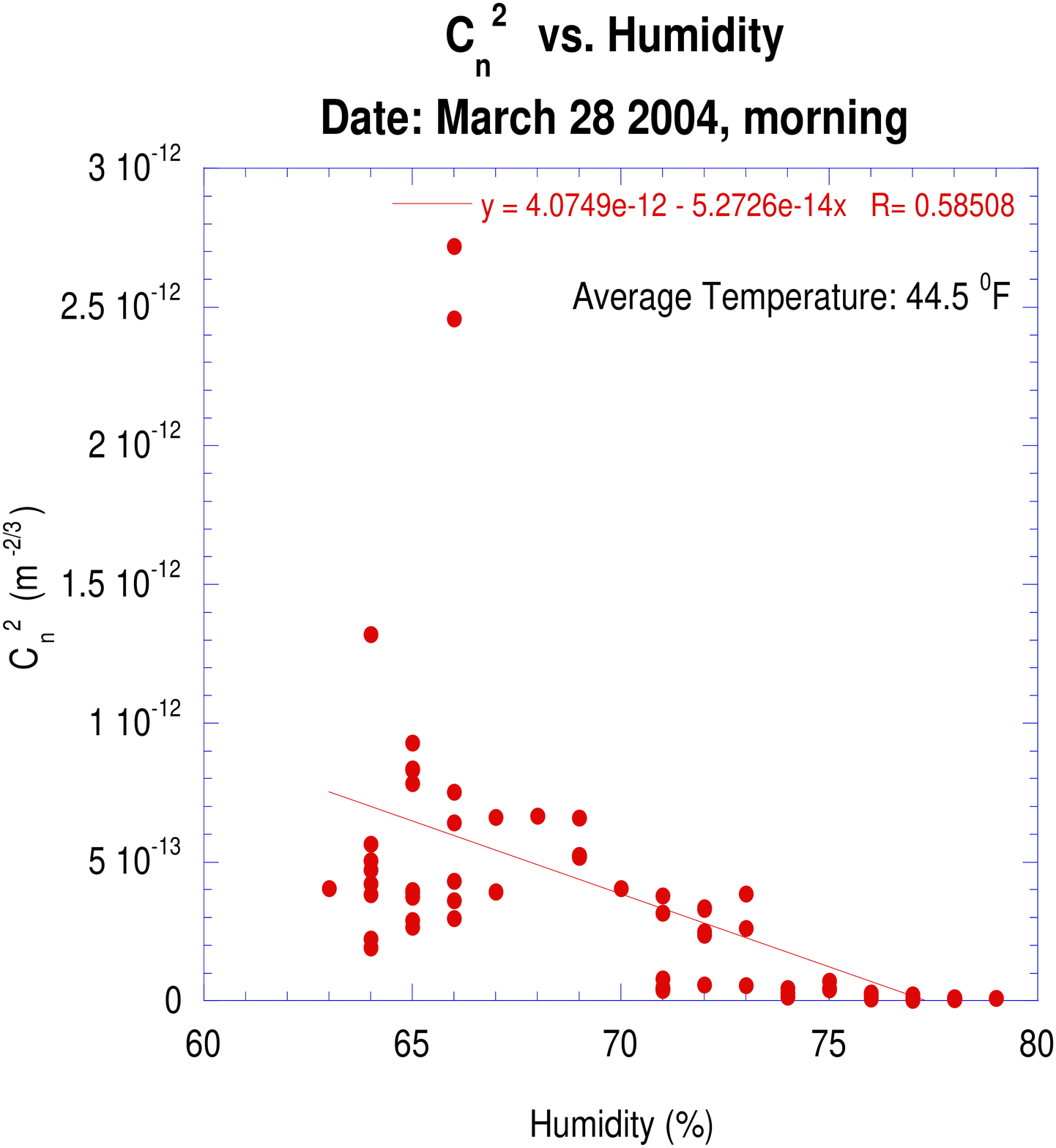}
\includegraphics[height=0.4\textwidth]{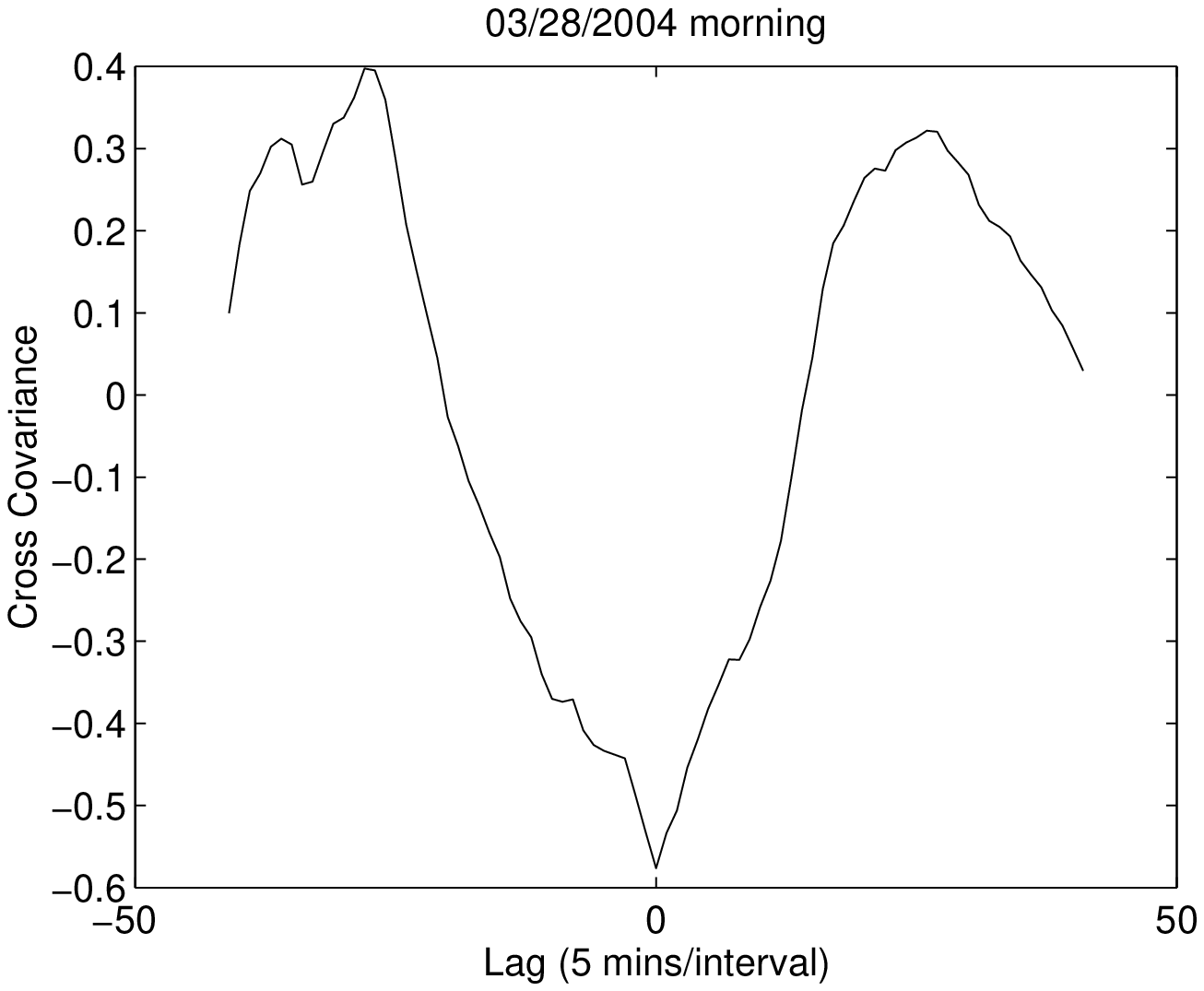}
\end{center}
\caption{\label{03282004} Mar 28 2004, Morning}
\end{figure}

\begin{figure}
\begin{center}
\includegraphics[height=0.44\textwidth]{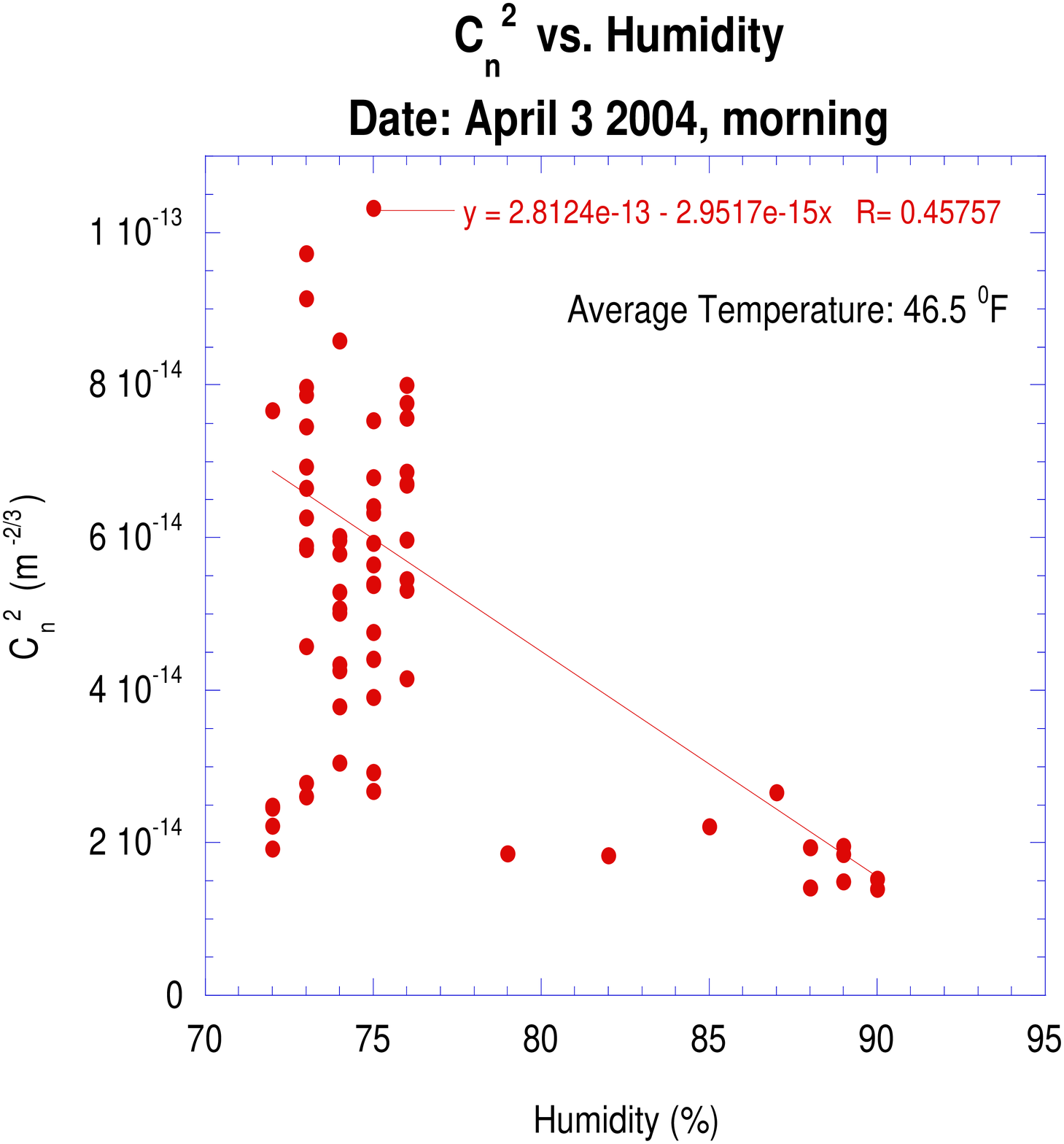} 
\includegraphics[height=0.4\textwidth]{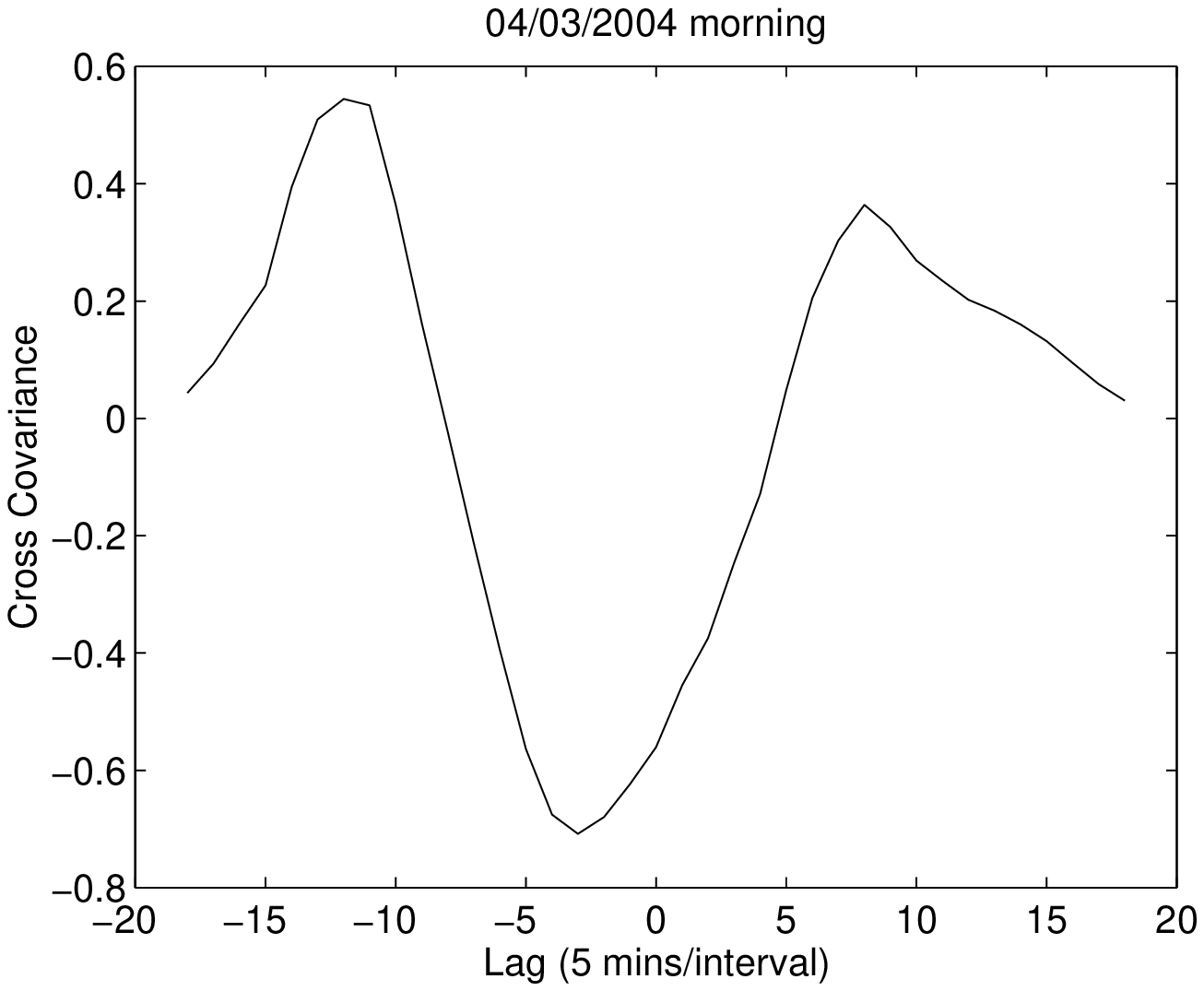}
\end{center}
\caption{\label{04032004m} Apr 3 2004 Morning}
\end{figure}

\begin{figure}
\begin{center}
\includegraphics[height=0.44\textwidth]{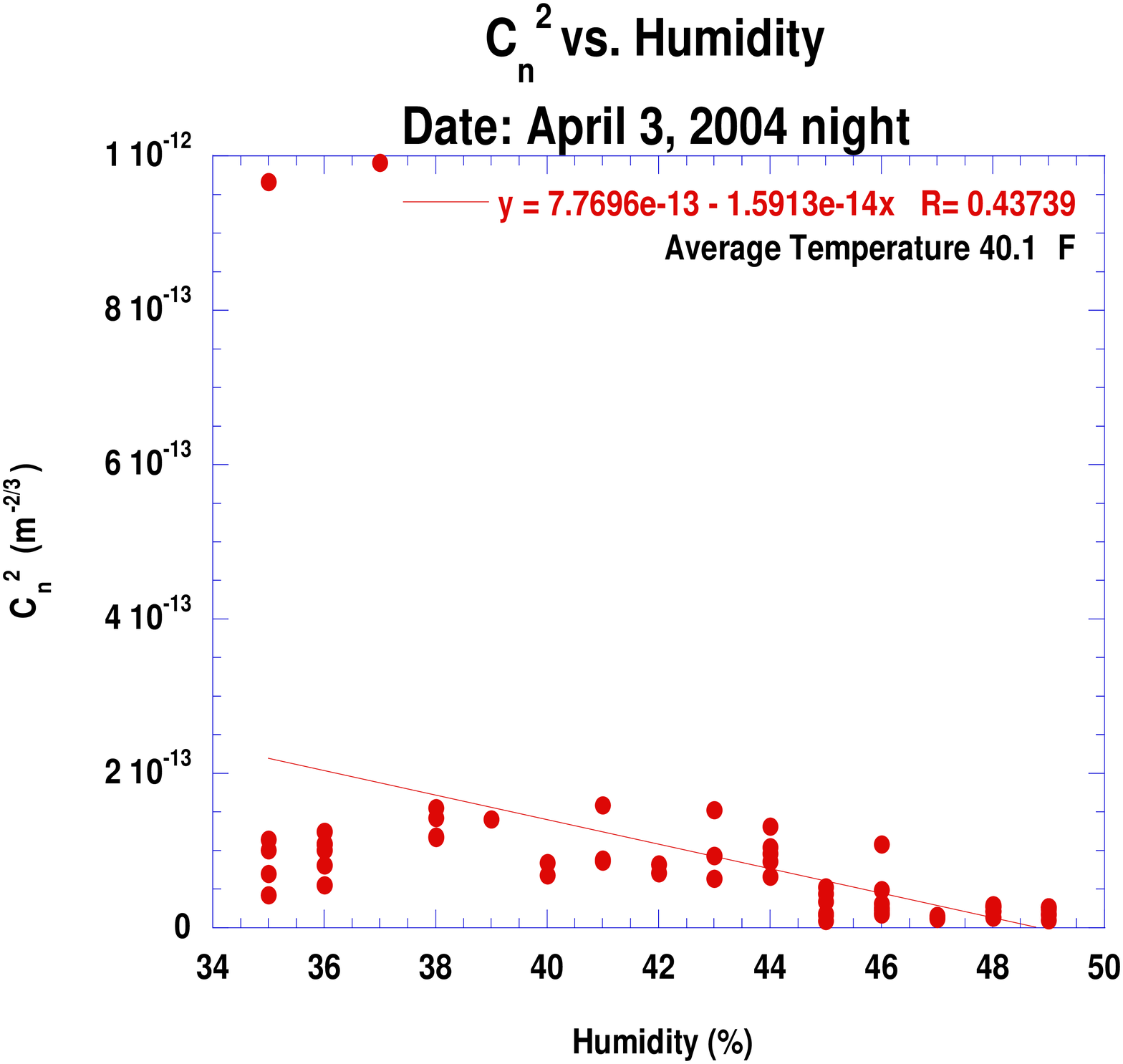}
\includegraphics[height=0.4\textwidth]{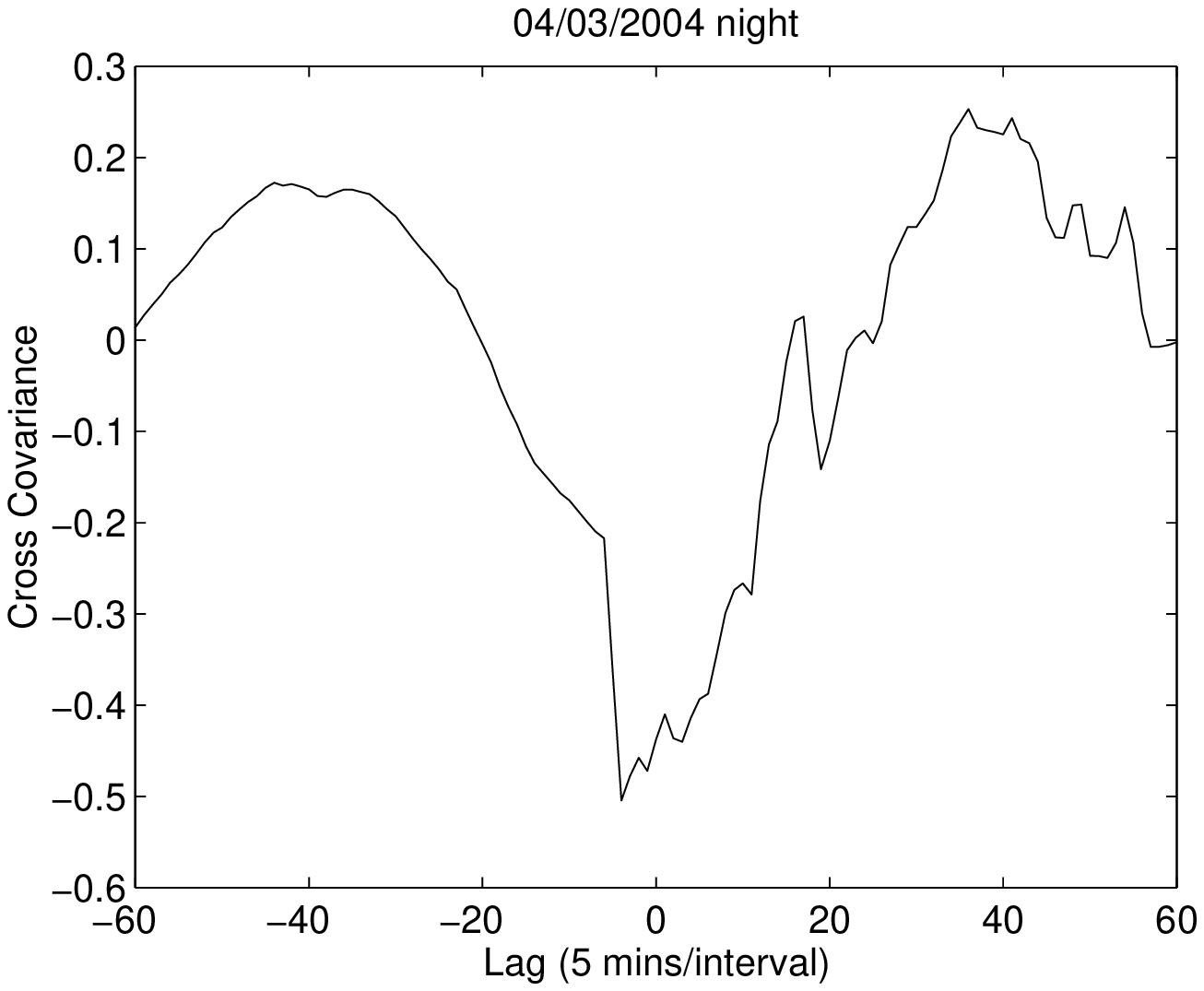}
\end{center}
\caption{\label{04032004n} Apr 3 2004, Night}
\end{figure}

As can be seen from Figures (\ref{11032003} - \ref{04032004n}), the $C_n^2$ against humidity correlograms all evidence a negative gradient.  The tightness of the correlation is better examined in terms of the cross covariance.  The results are normalized such that the value at zero time lag was unity for identically varying data.

\subsection{Comments on Figures (\ref{11032003} - \ref{04032004n})}

\begin{itemize}
\item[Fig.(\ref{11032003})] The correlogram shows an approximately even dispersion along the length of the best fit trendline.  The cross covariance lacks symmetry.
\item[Fig.(\ref{11092003})] The correlogram shows the tightest correlation between the data series of all the plots and the cross covariance is quite symmetric, although highly structured.
\item[Fig.(\ref{11102003})] The correlogram is fairly even, with some larger dispersion possibly occuring around 70\% of humidity.  The cross covariance is symmetric.
\item[Fig.(\ref{02022004})] The correlogram is evenly dispersed and the cross covariance is symmetric although spread.
\item[Fig.(\ref{03272004})] A greater dispersion is seen between 72\% and 74\% humidity than along the rest of the trendline, although in terms of magnitude it is not very large.  The cross correlation is highly asymmetric with the minimum offset from the zero lag position.
\item[Fig.(\ref{03282004})] A greater dispersion is seen below 67\% humidity than along the rest of the trendline (again the magnitude is not larger than the other plots).
\item[Fig.(\ref{04032004m})] A large cluster of weakly correlated points are seen below 77\% humidity.  The cross covariance minimum is offset from the zero lag position but is otherwise reasonably symmetric. 
\item[Fig.(\ref{04032004n})] The correlogram shows a reasonably good correlation between the data series, although the cross covariance shows less symmetry than might be expected from the correlogram.
\end{itemize}

\subsection{Covariance at zero lag, $C_{UV} (\delta t = 0)$}
\begin{figure}
\begin{center}
\begin{tabular}{c}
\includegraphics[height=0.45\textwidth]{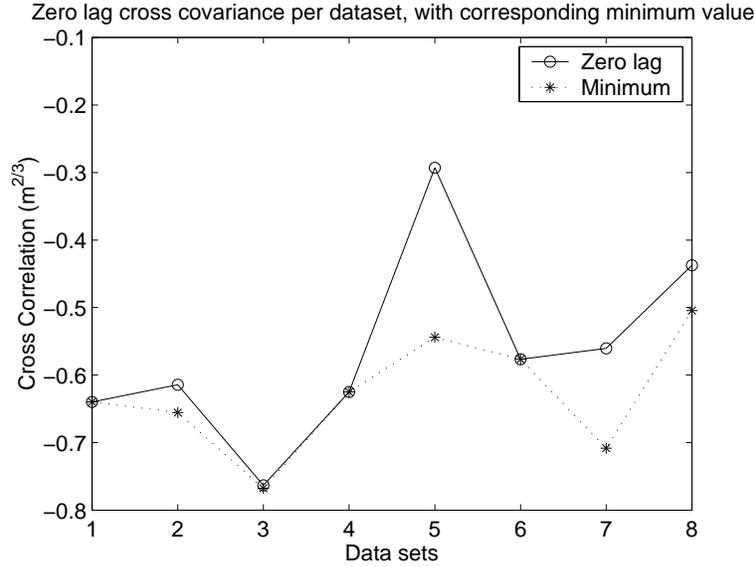}
\end{tabular}
\end{center}
\caption{\label{fig:xcov_zerolag} Cross covariance at zero lag.  The data set numbers are defined in Table (\ref{tbl:xcov_zerolag}).}
\end{figure}

The $C_{UV}$ at zero lag (i.e. when both $C_n^2$ and humidity data sets are totally overlapping) are given in Figure (\ref{fig:xcov_zerolag}) and Table (\ref{tbl:xcov_zerolag}).  The evidence for a negative correlation of humidity with $C_n^2$ is extremely strong; where the minimum cross covariances are different to the zero lag value, there is a time lag offset of no more than 25 minutes (equal to 5 sample points).  Some of the offset error is possibly due to a timing mismatch between the clocks used for the DP+ and the LOA-004 instruments; even without accounting for this, the $C_{UV} (\delta t = 0)$ is still strongly negative.

\begin{table}[!htp]
\begin{center}
\begin{tabular}{|c|l|c|} \hline
Data set number         & Date [mm/dd/yyyy]     & Zero timelag cross covariance \\ \hline
1                       & 11/03/2003 night      & -0.6397 \\
2                       & 11/09/2003 morning    & -0.6144 \\
3                       & 11/10/2003 morning    & -0.7632 \\
4                       & 02/02/2004 morning    & -0.6251 \\
5                       & 03/27/2004 morning    & -0.2930 \\
6                       & 03/28/2004 morning    & -0.5764 \\
7                       & 04/03/2004 morning    & -0.5604 \\
8                       & 04/03/2004 night      & -0.4374 \\ \hline
\end{tabular}
\end{center}
\caption{\label{tbl:xcov_zerolag} The cross correlation datasets.}
\end{table}

The cross covariance method has provided unequivocal measures that the humidity and $C_n^2$ functions are negatively correlated.

\section{CONCLUSIONS}
 
Using empirical data, we have conclusively demonstrated that a strong negative correlation exists between the humidity and $C_n^2$ readings from experimental runs at the Naval Research Laboratory's Chesapeake Bay Detachment, for path lengths of about 100-m with relatively constant pressure, temperature and windspeed.  On the basis of this we suggest that $C_n^2$ is an inverse function of humidity in the absence of solar insolation at coastal sites.

We are currently in the process of taking equivalent data at UPR-Mayag\"{u}ez and we expect to obtain measurements in the much less humid environment of New Mexico.  With the availability of more data, we will be able to ascertain in a quantitative fashion the humidity contribution to $C_n^2$.  We anticipate that a much deeper understanding of $C_n^2$ will be found from analysis of the complete data obtained under these extrememly varied ambient conditions. 

\acknowledgments     
MPJLC would like to thank Haedeh Nazari and Erick Roura for valuable discussions.



\bibliography{bib}   
\bibliographystyle{spiebib}   

\end{document}